\documentstyle[epsf]{depart}
\begin{document}
\baselineskip20pt
\begin{centering}  {\bf On the gravitomagnetic clock effect}\\
\vspace{28pt}
\end{centering} 

\noindent {\bf Bahram Mashhoon$^a,\;\;$Lorenzo Iorio$^b,\;\;$Herbert Lichtenegger$^c$}\\
\vspace{.25in}

\noindent$^a${\small Department of Physics and Astronomy, University of Missouri-Columbia,\\ Columbia, Missouri 65211,
USA\\
$^b$Dipartimento di Fisica dell' Universit\`{a} di Bari, via Amendola 173, 70126, Bari, Italy\\
$^c$Institut f\"{u}r Weltraumforschung, \"{O}sterreichische Akademie der Wissenschaften,\\ A-8042 Graz, Austria}

\vspace{.40in}

\begin{centering} {ABSTRACT}\\
\end{centering}

\vspace{.25in}

General relativity predicts that two freely counter-revolving test particles in the exterior field of a central
rotating mass take different periods of time to complete the same full orbit; this time difference leads to the
gravitomagnetic clock effect.  The effect has been derived for circular equatorial orbits; moreover, it has been
extended via azimuthal closure to spherical orbits around a slowly rotating mass.  In this work, a general formula is
derived for the main gravitomagnetic clock effect in the case of slow motion  along an arbitrary {\it elliptical} orbit
in the exterior field of a slowly rotating mass.  Some of the implications of this result are briefly discussed.\\

\newpage

\noindent 1.  Introduction

The gravitomagnetic clock effect is basically a reflection of the fact that according to general relativity there is a
special temporal structure around a rotating mass.  More specifically, let us consider circular geodesics in the
equatorial plane of a Kerr black hole of mass $M$ and angular momentum $J$.  Let
$t_{+}\:(t_{-})$ be the period of prograde (retrograde) motion along such an orbit according to asymptotically static
inertial observers; then, $t_{+}-t_{-} = 4\pi J/(Mc^2)$.  Moreover, let
$\tau_{\pm}$ represent the corresponding proper periods according to comoving clocks; then $\tau_{+}-\tau_{-}\approx
4\pi J/(Mc^2)$ for orbits with radius $r>>2GM/c^2$.  It is remarkable that at $O(c^{-2})$ this result is independent of
the gravitational constant
$G$ and the radius of the orbit [1,2].  Various theoretical aspects of this effect have been investigated [3-10].  On
the observational side, the possibility of its detection has been considered by a number of authors [11-17].  

The general relativistic calculation of the gravitomagnetic clock effect for a general orbit is quite complicated. 
Other than equatorial circular orbits, only spherical orbits have been considered thus far to first order in the
rotational perturbation within the post-Schwarzschild approximation scheme [3].  To ameliorate this situation, we
recently showed that it is possible to recover the main general relativistic results at the lowest order, i.e.
$O(c^{-2})$, by considering the linear approximation of general relativity for the exterior geometry of a slowly
rotating mass [18].  The equations of motion of a test particle in such a stationary field reduce, in the slow-motion
approximation, to an equation of the Lorentz form, $d{\bf v}/dt = {\bf E}_g +
\mbox{\boldmath${\cal E}$}_g + ({\bf v}/c) \times {\bf B}_g$, where ${\bf E}_g = -GM{\bf r}/r^3$ is the Newtonian
gravitoelectric field, $\mbox{\boldmath${\cal E}$}_g$ is the post-Newtonian gravitoelectric field and ${\bf B}_g$, 

\begin{equation}  {\bf B}_g = \frac{2G}{cr^5}\left[{\bf J}r^2 - 3({\bf J}\cdot{\bf r}){\bf r}\right]\;\;,
\end{equation}

\noindent is the post-Newtonian gravitomagnetic field (see [19, 20]
 for reviews of gravitoelectromagnetism). 
It should be noted that there is a certain arbitrariness in the
 definition of the gravitoelectromagnetic field depending on how the
 analogy with electrodynamics is developed; in particular, the present
 treatment is different from the development in [20]. In this
 connection, we remark in passing that the approach that is closest to
 electromagnetism is based on the gravitational Larmor theorem (cf.\
 [3,7,20] and references therein).
The explicit form of
$\mbox{\boldmath${\cal E}$}_g$ can be simply derived using the isotropic form of the Schwarzschild metric along the
lines of equation (10) of [20].  To $O(c^{-2}), \mbox{\boldmath${\cal E}$}_g$ is given by $c^2\mbox{\boldmath${\cal
E}$}_g =(v^2-4\Phi)\:{\bf E}_g-4({\bf E}_g\cdot {\bf v}){\bf v}$, where $\Phi=GM/r$ is the Newtonian potential.  Note
that at a given point ${\bf r}$ along the orbit, reversing the direction of motion, ${\bf v}\rightarrow - {\bf v}$,
leaves $\mbox{\boldmath${\cal E}$}_g$ invariant while the gravitomagnetic perturbation changes sign.  The contribution
of $\mbox{\boldmath${\cal E}$}_g$ to the orbital period is thus independent of the sense of orbital revolution.  In
fact, the orbital perturbation due to
$\mbox{\boldmath${\cal E}$}_g$ is planar and leads to the Einstein
 pericenter precession. 
Calculating the period of motion in this case via azimuthal closure (i.e. the time that it takes for the particle to go
from
$\phi=\phi_{0}$ to $\phi=\phi_{0} + 2\pi$ or vice versa), one finds that there is no clock effect when otherwise
identical clockwise and counterclockwise orbital periods are compared.  Hence, in the calculation of the clock
effect to $O(c^{-2})$, the post-Newtonian gravitoelectric perturbation due to the mass $M$ may be neglected. 
Therefore, for the sake of simplicity we essentially ignore
$\mbox{\boldmath${\cal E}$}_g$ in the following sections but return to it in the Appendix, where the gravitoelectric
contribution to the orbital period is explicitly given. 

The gravitomagnetic clock effect has not yet been detected observationally; therefore,
$O(c^{-2})$ terms are the only ones that are of current interest.  At this level of approximation, we derive the
general effect in this paper for an eccentric orbit.  Though our approach is general, we have in mind for specific
applications the motion of artificial satellites around the Earth.  In the absence of the gravitomagnetic force, the
orbit would ideally be a Keplerian ellipse; therefore, we treat the gravitomagnetic force as a linear perturbation of
the dominant Newtonian gravitoelectric force.  Since such a perturbation cannot be ``turned off'' for motion around the
Earth, the elements of the unperturbed elliptical orbit play the role of subsidiary variables in our treatment:  They
are not all directly measurable.  The unperturbed orbital plane, however, is given by the instantaneous plane of the
osculating ellipse when the observations begin by definition at $t=0$.  

In Section 2 we derive the general form of the perturbed orbit within our approximation scheme.  The ``period'' of the
motion is calculated in Section 3 using the notion of azimuthal closure.  Section 4 presents the general formula for
the gravitomagnetic clock effect to $O(c^{-2})$ as well as a discussion of some of its main implications.  

\vspace{.25in}

\noindent 2.  Eccentric orbit

Let us consider a solution of the equations of motion

\begin{equation}
\frac{d^2{\bf r}}{dt^2} = -\frac{GM{\bf r}}{r^3} + \frac{2G}{c^2r^5}\left[3({\bf r}\cdot{\bf J}){\bf r}
\times {\bf v} + r^2{\bf v}\times{\bf J}\right]\;\;,
\end{equation}

\noindent which would correspond, in the absence of the perturbation, to the unperturbed elliptical orbit as in Figure
1.  To simplify the analysis, we introduce the ``inertial'' coordinate system (X, Y, Z) adapted to the unperturbed
orbit (as in Fig. 1) such that the unperturbed orbit is given by $X =
\rho_0\:{\rm cos}\:\varphi, Y = \rho_0\:{\rm sin}\:\varphi$ and $Z=0$, where

\begin{equation}
\rho_0 = \frac{a_0(1-e_0^2)}{1+e_0\:{\rm cos}\:\hat{v}}\;\;,\;\;\omega_{0}t=(\hat{u}-e_0\:{\rm
sin}\:\hat{u})-(\hat{u}-e_0\:{\rm sin}\:\hat{u})_{t=0}\;\;.
\end{equation}

\noindent Here $a_0, e_0, \hat{v}=\varphi-g_0$ and $\hat{u}$ are respectively the semimajor axis, eccentricity, true
anomaly and the eccentric anomaly of the unperturbed orbit; moreover, the angle $g_0$ is the argument of pericenter
as depicted in Figure 1.  The Keplerian frequency and the period of this orbit are given by $\omega_0 =
(GM/a_0^3)^{1/2}$ and
$T_0=2\pi/\omega_{0}$, respectively.  Let us note for future reference that
$\rho_0 = a_0(1-e_0\:{\rm cos}\:\hat{u})$ and
$a_0(1-e^2_0)\dot{\rho}_0=e_0L_0\:{\rm sin}\:\hat{v}$, where an overdot represents differentiation with respect to time
$t$ and
$L_0=\sqrt{GMa_0(1-e_0^2)}$ is the specific angular momentum of the unperturbed orbit.

To write the equations of motion in a convenient form, we express equation (2) in terms of the $(X, Y, Z)$ coordinate
system.  It proves useful to introduce the cylindrical coordinate system $(\rho,
\varphi, Z)$ as well, where $X = \rho\:{\rm cos}\:\varphi$ and
$Y=\rho\:{\rm sin}\:\varphi$.  In the ``inertial'' $(X, Y, Z)$ system, the equations of motion can be expressed as
$\ddot{X} +GMX/\rho^3 = F_X$ and similarly for the $Y$ and $Z$ components.  The advantage of this system is that along
the orbit $Z$ can be treated only to first order in the perturbation; therefore, the radial position of the test
particle is in effect given by $\rho=(X^2 + Y^2)^{1/2}$ and the perturbing acceleration $(F_X, F_Y, F_Z)$ is simply
evaluated along the unperturbed orbit.  The equations of motion can then be written in the $(\rho, \varphi, Z)$
system as 

\begin{eqnarray}
\ddot{\rho}-\rho\dot{\varphi}^2+\frac{GM}{\rho^2}&=&F_{\rho}\;\;,\\
\rho\ddot{\varphi}+2\dot{\rho}\dot{\varphi} &=& F_{\varphi}\;\;,\\
\ddot{Z} + \frac{GM}{\rho^3}Z &=& F_Z\;\;.
\end{eqnarray}

\noindent Here $F_{\rho}=F_X\:{\rm cos}\:\varphi + F_Y\:{\rm sin}\:\varphi$ and
$F_{\varphi}=-F_X\:{\rm sin}\:\varphi + F_Y\:{\rm cos}\:\varphi$. The perturbing accelerations are given by $F_{\rho} =
\epsilon L_0/\rho_0^4, F_{\varphi} = -\epsilon\dot{\rho}_0/\rho_0^3$ and

\begin{equation} F_Z = \frac{2L_0GJ\:{\rm sin}\:i}{c^2\rho_0^3 a_0(1-e_0^2)}\left[ 2(1+e_0\:{\rm cos}\:{\hat v})\:{\rm
sin}\:\varphi + e_0\:{\rm sin}\:{\hat v}\:{\rm cos}\:\varphi\right] \;\;,
\end{equation}

\noindent where $\epsilon =2GJ\:{\rm cos}\:i/c^2$ and ${\hat v}=\varphi-g_0$.

To solve the equations of motion to first order in $\epsilon$, we start with equation (5) and write it in the form
$d(\rho^2\dot{\varphi})/dt=\rho F_{\varphi}$, which expresses the rate of change of the
$Z$-component of the orbital angular momentum.  This equation can be easily integrated and the result is
$\rho^2\dot{\varphi}=C+\epsilon/\rho_0$, where $C$ is a constant of integration that reduces to
$L_0$ in the absence of the perturbation.  In fact, we can write $C = L_0(1+\epsilon\lambda)$, which defines the
constant
$\lambda$.  We can now use the result of the integration of equation (5) to write equation (4) as a differential
equation for
$u=1/\rho$ as a function of
$\varphi$ in the standard manner.  The result is 

\begin{equation}
\frac{d^2u}{d\varphi^2}+u = \frac{GM}{C^2}-\epsilon\:\frac{(3+e^2_0)+4e_0\:{\rm
cos}\:(\varphi-g_0)}{L_0a^2_0(1-e^2_0)^2}\;\;.
\end{equation}

\noindent The general solution of equation (8) is given by 

\begin{eqnarray} u =
\frac{GM}{C^2}&-&\epsilon\:\frac{3+e_0^2}{L_0a_0^2(1-e_0^2)^2}-2\epsilon\:\frac{e_0\varphi\:{\rm
sin}\:(\varphi-g_0)}{L_0a_0^2(1-e_0^2)^2}\nonumber\\ &+&K\:{\rm cos}\:(\varphi - g_0)+K^{\prime}\:{\rm
sin}\:(\varphi-g_0)\;\;,
\end{eqnarray}

\noindent where $K$ and $K^{\prime}$ are integration constants.  Let us note that equation (9) should reduce to the
unperturbed ellipse given in equation (3) for $\epsilon=0$.  It follows from this requirement that 

\begin{equation} K=\frac{e_0}{a_0(1-e_0^2)}+\epsilon k\;\;\;,\;\;\;K^{\prime}=\epsilon k^{\prime}\;\;,
\end{equation}

\noindent where $k$ and $k^{\prime}$ are constants.  It is possible to write equation (9) in the form

\begin{equation} u=\frac{1+e\:{\rm cos}\:[(1+\delta)\varphi-g]}{a(1-e^2)}\;\;,
\end{equation}

\noindent where $\delta$ is a small dimensionless perturbation parameter defined by 

\begin{equation}
\delta=\frac{4\omega J\:{\rm cos}\:i}{Mc^2(1-e^2)^{3/2}}\;\;.
\end{equation}

\noindent Here $\omega=(GM/a^3)^{1/2}$ and the period corresponding to this ``Keplerian'' frequency is
$T=2\pi/\omega$.  Equation (11) contains three new constant orbital elements $a, e$ and $g$ defined by

\begin{equation} a=a_0(1+\epsilon\alpha),\;e=e_0(1+\epsilon\eta),\;g=g_0+\epsilon\beta k^{\prime}\;\;.
\end{equation}

\noindent In a realistic situation, such as the motion of a satellite around the Earth, the gravitomagnetic
perturbation caused by the rotation of the Earth cannot be turned off.  Therefore,
$a_0, e_0$ and $g_0$ should be thought of as auxiliary constants that are not directly measurable.  On the other hand,
by comparing the observed orbit with equation (11), one can determine the orbital parameters $a, e$ and $g$.  Using the
definitions (13) in equation (11) and comparing the result with equation (9), we arrive at the following expressions
for $\alpha, \beta$ and $\eta$:

\begin{eqnarray}
\alpha&=&\frac{1+e_0^2}{1-e_0^2}\left[ 2\lambda+\frac{3+e_0^2}{L_0a_0(1-e_0^2)} \right] + 2e_0a_0k\;\;,\\
\beta&=&\frac{a_0}{e_0}(1-e_0^2)\;\;,\\
\eta&=&\frac{1-e^2_0}{1+e_0^2}(\alpha+\beta k)\;\;,
\end{eqnarray}

\noindent where $\lambda$ is given by the relation $C=L_0(1+\epsilon\lambda)$ and $k$ is defined in equation (10).

The variation of $\varphi$ with time $t$ is given by $\dot{\varphi}=Cu^2+\epsilon u^3$, where
$u(\varphi)$ is given by equation (11).  Here $C$ is different from
$L=\sqrt{GMa(1-e^2)}$; in fact, using equations (14) - (16) we find that 

\begin{equation}  C=L-\frac{1}{2}\epsilon\:\frac{3+e^2}{a(1-e^2)}\;\;.
\end{equation}

To solve equation (6) for $Z$, we employ the approach described in [21]:  Let $Z=\rho H(\varphi)$; then, using
equations (4) and (5) we can write equation (6) as 

\begin{equation}
\rho\dot{\varphi}^2\left( \frac{d^2H}{d\varphi^2}+H\right) +F_{\rho}H+F_{\varphi}\frac{dH}{d\varphi}=F_Z\;\;.
\end{equation}

\noindent Since $H=O(c^{-2})$ and
$F_{\rho}H+F_{\varphi}\:{dH}/{d\varphi}=O(c^{-4})$, equation (18) implies that 

\begin{equation}
\frac{d^2H}{d\varphi^2} +H =\delta^{\prime}\left[\:{\rm sin}\:\varphi+e\:{\rm sin}\:\varphi\:{\rm
cos}\:(\varphi-g)+{1\over 2} e\:{\rm cos}\:\varphi\:{\rm sin}\:(\varphi-g)\right]\;\;,
\end{equation}

\noindent where $\delta^{\prime}$ is another small dimensionless perturbation parameter defined by 

\begin{equation}
\delta^{\prime}=\frac{4\omega J\:{\rm sin}\:i}{Mc^2(1-e^2)^{3/2}}\;\;.
\end{equation}

We assume that when observations begin at $t=0, \varphi=\varphi_0$ and the instantaneous orbital plane coincides with
the plane of  the unperturbed orbit depicted in Fig. 1.  That is, at $\varphi =
\varphi_0$ we have $H=0$ and $dH/d\varphi=0$ since both $Z$ and
${\dot Z}$ vanish.  With these initial conditions, the general solution of equation (19) is given by

\begin{eqnarray} H &=& \frac{\delta^{\prime}}{4}\left[\:{\rm sin}\:\varphi+\:{\rm
sin}\:(\varphi-2\varphi_0)-2(\varphi-\varphi_0)\:{\rm cos}\:\varphi+e\:{\rm
sin}\:(\varphi+\varphi_0-g)\right.\nonumber\\ &+&e\:{\rm sin}\:(\varphi-\varphi_0)\:{\rm cos}\:(2\varphi_0-g)-e\:{\rm
cos}\:(\varphi-\varphi_0)\:{\rm sin}\:g-e\:{\rm sin}\:(2\varphi-g)\nonumber\\ &+&e\:{\rm sin}\:g\left.\right]\;\;.
\end{eqnarray}

\noindent This completes the construction of the orbit that is given by $X=\rho\;{\rm
cos}\:\varphi\;\;,\;\;Y=\rho\;{\rm sin}\:\varphi$ and $Z=\rho H(\varphi)$.  Here
$\rho=1/u(\varphi)$, where $u(\varphi)$ is given by equation (11), and $H(\varphi)$ is given by equation (21).  From
the standpoint of observers in the basic $(x, y, z)$ coordinate system, the orbit is characterized by the elements $a,
e, i, g, \Omega$ and
$\varphi_0$, where $\Omega$ is the longitude of the ascending node; indeed these six orbital elements can be determined
in principle from the position and velocity of the satellite at $t=0$.  

It is straightforward to show that the orbit thus constructed undergoes Lense-Thirring precession.  To this end, one
could start from the equation of motion (2) and study the average behavior of the Runge-Lenz vector and the orbital
angular momentum vector as in [22].  Alternatively, one could study explicitly the orbit derived here.  For instance
let us note that for $i=0$, the orbit is planar and the perigee occurs at
$\varphi=(1-\delta)(g+2n\pi), n=0, 1, 2, ...$.  Thus the orbit precesses in the retrograde direction at the rate of
$2\pi\delta/T=4GJ/[c^2a^3(1-e^2)^{3/2}]$, as expected [19].

\vspace{.25in}

\noindent 3.  Azimuthal closure

We assume that orbital observations are usually performed with respect to the fundamental frame $(x, y, z)$.  The
transformation between the $(X, Y, Z)$ and $(x, y, z)$ frames is given by the rotation matrix

\begin{equation}  
\left[ \begin{array}{c} x \\ y \\ z\end{array}\right]\;\;=
\left[ \begin{array}{ccc} {\rm cos}\:\Omega & -{\rm sin}\:\Omega\:{\rm cos}\:i & {\rm sin}\:\Omega\:{\rm sin}\:i\\ {\rm
sin}\:\Omega & {\rm cos}\:\Omega\:{\rm cos}\:i & -{\rm cos}\:\Omega\:{\rm sin}\:i\\ 0 & {\rm sin}\:i & {\rm
cos}\:i\end{array}\right]\;\;
\left[ \begin{array}{c} X \\ Y \\ Z\end{array}\right]\;\;.
\end{equation}

\noindent In the basic $(x, y, z)$ frame, let us introduce polar coordinates via
$x\;=\;r\:{\rm sin}\:\theta\:{\rm cos}\:\phi$, $y=r\:{\rm sin}\:\theta\:{\rm sin}\:\phi$ and $z=r\:{\rm cos}\:\theta$. 
Let us note that to $O(c^{-2}), r = \rho$ along the orbit.  We are interested in the time
${\cal T}$ that the particle would take to go from $\phi_0$ at $t=0$ to
$\phi_0+2\pi$ at $t={\cal T}$. The orbit is not closed in space; therefore, we seek the period for azimuthal closure
with respect to the basic coordinate system that is used for observations.  Since

\begin{equation}  \tan\phi=\frac{y}{x}=\frac{{\rm sin}\:\Omega\:{\rm cos}\:\varphi+{\rm cos}\:\Omega\:{\rm
cos}\:i\:{\rm sin}\:\varphi-\:{\rm cos}\:\Omega\:{\rm sin}\:i\:H(\varphi)}{{\rm cos}\:\Omega\:{\rm cos}\:\varphi-\:{\rm
sin}\:\Omega\:{\rm cos}\:i\:{\rm sin}\:\varphi+{\rm sin}\:\Omega\:{\rm sin}\:i\:H(\varphi)}\;\;,
\end{equation}

\noindent we evaluate $\tan\phi_0$ from equation (23) at $\varphi=\varphi_0$ (corresponding to
$t=0$) and set it equal to ${\rm tan}\:(\phi_0+2\pi)$ evaluated from equation (23) at
$\varphi=\varphi_{\cal T}$ (corresponding to
$t={\cal T}$).  This simply implies, after some algebra, that for ${\rm cos}\;i\neq 0$,

\begin{equation}  {\rm sin}\:(\varphi_{\cal T}-\varphi_0)=[{\rm cos}\:\varphi_0\:H(\varphi_{\cal T})-{\rm
cos}\:\varphi_{\cal T}\:H(\varphi_0)]\:{\rm tan}\:i\;\;.
\end{equation}

\noindent Let us suppose that $\varphi_{\cal T} = \varphi_0+2\pi+\Delta$, where $\Delta$ is yet another small quantity;
then, it follows from equation (24) that $\Delta = -\pi\delta^{\prime}\:{\rm cos}^2\:\varphi_0\;{\rm tan}\:i$, since
$H(\varphi_0)=0$ by assumption and
$H(\varphi_{\cal T}) = -\pi\delta^{\prime}\:{\rm cos}\:\varphi_0$ according to equation (21).  To ensure that
$|\Delta|<<1$, the inclination angle $i$ must not be near $\pi/2$ since ${\rm tan}\:(\pi/2)=\infty$.  Thus azimuthal
closure in $\phi$ corresponds to
$\varphi : \varphi_0\rightarrow\varphi_0+2\pi+\Delta$.  

It now remains to integrate $\dot{\varphi}=Cu^2+\epsilon u^3$ over the interval $(\varphi_0, \varphi_{\cal T})$ in
order to determine
${\cal T}$, i.e.

\begin{equation} {\cal T} = \int_{\varphi_0}^{\varphi_{\cal T}}\frac{d\varphi}{Cu^2+\epsilon u^3}\;\;,
\end{equation}

\noindent where $u(\varphi)$ is given by equation (11).  To first order in $\Delta$, equation (25) can be written as 

\begin{equation} {\cal T} = \int_{\varphi_0}^{\varphi_0+2\pi}\frac{d\varphi}{Cu^2+\epsilon u^3} +
\frac{\Delta(1-e^2)^{3/2}}{\omega[1+e\:{\rm cos}(\varphi_0 - g)]^2}\;\;.
\end{equation}

\noindent Moreover, equations (11) and (17) can be used to show that 

\begin{eqnarray} (Cu^2+\epsilon u^3)^{-1} &=& \frac{(1-e^2)^{3/2}}{\omega[1+e\:{\rm
cos}\:(\varphi-g)]^2}\left[1+2\delta\frac{e\varphi\:{\rm sin}\:(\varphi-g)}{1+e\:{\rm cos}\:(\varphi-g)} \right.
\nonumber\\ &+&\left.\frac{1}{2}\epsilon\:\frac{1-2e\:{\rm cos}\:(\varphi-g)+e^2}{\omega a^3(1-e^2)^{3/2}}\right]\;\;.
\end{eqnarray}

\noindent Substituting this expression for the integrand in equation (26), we obtain ${\cal T}$ in terms of three basic
integrals.  To compute these,we recall that 

\begin{equation}
\int^{\zeta_0+2\pi}_{\zeta_0}\frac{d\zeta}{1+e\:{\rm cos}\:\zeta} =
\frac{2\pi}{(1-e^2)^{1/2}}\;\;,\;\;\int^{\zeta_0+2\pi}_{\zeta_0}\frac{d\zeta}{(1+e\:{\rm cos}\:\zeta)^2}
=\frac{2\pi}{(1-e^2)^{3/2}}\;\;;
\end{equation}

\noindent moreover, one can show via integration by parts that 

\begin{equation}
\int^{\zeta_0 +2\pi}_{\zeta_0}\frac{(\zeta+g)e\:{\rm sin}\:\zeta d\zeta}{(1+e\:{\rm cos}\:\zeta)^3} =
\frac{\pi}{(1+e\:{\rm cos}\:\zeta_0)^2} - \frac{\pi}{(1-e^2)^{3/2}}\;\;.
\end{equation}

\noindent The result can be written as 

\begin{equation}
\frac{{\cal T}}{T} = 1 + \frac{GJ\:{\rm cos}\:i}{c^2a^3\omega}\left\{-\frac{3}{\sqrt{1-e^2}} + \frac{4-2\:{\rm
cos}^2\:\varphi_0\:{\rm tan}^2\:i}{[1+e\:{\rm cos}\:(\varphi_0-g)]^2}\right\}\;\;.
\end{equation}

\noindent This expression reduces to our previous result [18] for a spherical orbit with
$e=0$.  We emphasize that equation (30) properly contains only the {\it gravitomagnetic} correction to the period
${\cal T}$ to $O(c^{-2})$, i.e. ${\cal T} = T(1+\Theta_{gm})$.  The corresponding gravitoelectric perturbation,
$\Theta_{ge}$, is given by equation (34) of the Appendix.  Therefore, the final result may be written as ${\cal
T}_{\pm}=T(1+\Theta_{ge}\pm \Theta_{gm})$, where the upper (lower) sign refers to a prograde (retrograde) orbit.  

\vspace{.25in}

\noindent 4. Discussion

The prograde orbit for which we have derived the expression for ${\cal T}$ in equation (30) is such that the particle
revolves around the source in the same sense as the proper rotation of the source (see Fig. 1).  Imagine now the
retrograde case such that at each point on the orbit the direction of the velocity vector is reversed.  If at the same
time the direction of ${\bf J}$ --- and hence
${\bf B}_g$ --- is reversed as well, then ${\bf E}_g + ({\bf v}/c)\times{\bf B}_g$ would be unchanged; therefore, the
particle would be subject to the same force as before and relation (30) would hold in this case as well.  It follows
that if in Fig. 1 the orbit is assumed to be retrograde, then equation (30) would hold except that the overall sign of
the perturbation term would be reversed.  Thus the general formula for the clock effect is 

\begin{equation} {\cal T}_+ - {\cal T}_- = 4\pi\frac{J\:{\rm cos}\:i}{Mc^2}\left\{-\frac{3}{\sqrt{1-e^2}} +
\frac{4-2\:{\rm cos}^2\:\varphi_0\:{\rm tan}^2\:i}{[1+e\:{\rm cos}\:(\varphi_0-g)]^2} 
\right\}\;\;.
\end{equation}

\noindent We expect that formula (31) would hold at
$O(c^{-2})$ level as well if ${\cal T}_+$ and ${\cal T}_-$ were replaced by the proper periods of spaceborne clocks on
pro- and retro-grade orbits, respectively. 

It is interesting to note the dependence of the gravitomagnetic clock effect on the eccentricity of the orbit; this is
illustrated in Figure 2 for an artificial satellite in orbit around the Earth.  Though equation (31) is formally valid
for any eccentricity, $e$ cannot be too close to unity; otherwise, the small dimensionless perturbation parameters
$\delta$ and $\delta^{\prime}$ given by equations (12) and (20), respectively, would no longer be small and our
perturbation analysis would break down. 

A remarkable feature of equation (31) is its topological character, i.e. its complete independence from the semimajor
axis of the orbit.  Moreover, the gravitational constant $G$ does not appear in equation (31).  For the Earth, $4\pi
J/(Mc^2)\sim 10^{-7}{\rm s}$; therefore, it may be possible to measure this effect in the near future [11-17].  

The clock effect vanishes for a polar orbit.  Equation (31) thus breaks down when the inclination angle
$i$ is sufficiently close to $\pi/2$.  This circumstance is related to the fact that for a spherical polar geodesic
orbit, the orbit precesses with the Lense-Thirring frequency $2GJ/(c^2a^3)$ in the same sense as the rotation of the
source.  Therefore, the azimuthal closure period is in this case $2\pi(2GJ/c^2a^3)^{-1}$.  This period is extremely
long compared to the orbital period; in fact, the corresponding ratio is $\sim 10^{10}$ for near-Earth orbits.  Thus in
our perturbative approach $i$ must be sufficiently different from $\pi/2$ such that $|\Delta|<<1$.  This fact is
reflected in the requirement that in equation (30) the relative gravitomagnetic perturbation in the period be much less
than unity.  

Finally, it is interesting to note that the clock effect (31) in general depends on the position of the particle on its
orbit at
$t=0$.  In fact, of the six orbital elements $(a, e, i, g, \Omega, \varphi_0)$, only $a$ and $\Omega$ are not
represented in equation (31) due to the topological character of the effect and the assumed axial symmetry of the
source, respectively.  

\vspace{.25in}

\newpage
\noindent Appendix

The general solution of the gravitoelectromagnetic equation of motion to $O(c^{-2})$ can be obtained via  a
superposition of the perturbations due to $\mbox{\boldmath${\cal E}$}_g$ and ${\bf B}_g$.  The latter have been the
main subject of this paper.  For the sake of completeness, we wish to indicate briefly in this Appendix the result of
the perturbation analysis for
$\mbox{\boldmath${\cal E}$}_g$.  The work described in Sections 2 and 3 can be repeated for $\mbox{\boldmath${\cal
E}$}_g$ as the perturbing acceleration.  Indicating the corresponding quantities by a tilde, the perturbing
accelerations are given by 

\begin{equation} {\tilde F}_{\rho} = \frac{G^2M^2}{c^2\rho_0^2}\left[ \frac{2}{\rho_0} + \frac{1-e^2_0+4e_0^2\:{\rm
sin}^2\:{\hat v}}{a_0(1-e_0^2)}\right]\;\;,
\end{equation}

\noindent ${\tilde F}_{\varphi} = -{\tilde\epsilon}{\dot\rho}_0/\rho^3_0$ and ${\tilde F}_{Z} = 0$, where
${\tilde\epsilon} = -4GML_0/c^2$.  The analysis of Section 2 can be followed in much the same way with
$\epsilon\rightarrow{\tilde\epsilon}$.  The result is $Z=0$ and an orbit of the form (11) with
$\delta\rightarrow{\tilde\delta}$, where 

\begin{equation}  {\tilde\delta} = -3\frac{GM}{c^2a(1-e^2)}\;\;,
\end{equation}

\noindent and ${\dot\varphi}={\tilde C}u^2 + {\tilde\epsilon}u^3$ such that ${\tilde C}$ is related to $L$ as in
equation (17) with
$\epsilon\rightarrow{\tilde\epsilon}$ and $3+e^2\rightarrow(5+e^2)/4$.  Equation (33) implies that the orbit undergoes
Einstein precession by $2\pi|{\tilde\delta}|$ radian per revolution.  The period of the orbit ${\tilde{\cal T}}$ can be
obtained from the integration of ${\dot\varphi} = {\tilde C}u^2+{\tilde\epsilon}u^3$ for
$\varphi:\varphi_0\rightarrow\varphi_0+2\pi$, which agrees with $\phi:\phi_0\rightarrow\phi_0+2\pi$ except when
$i=\pi/2$.  The integration can be performed as in Section 3 and the result is
${\tilde{\cal T}}=T(1+\Theta_{ge})$, where

\begin{equation}
\Theta_{ge} = \frac{3GM}{2c^2a}\left\{3-2\frac{\sqrt{1-e^2}}{\left[1+e\:{\rm cos}\:(\varphi_0-g)\right]^2}\right\}\;\;.
\end{equation}

\noindent For a circular orbit $e=0$, we find $\Theta_{ge}=3GM/(2c^2a)$.  This has a simple physical explanation in
terms of the isotropic radial coordinate employed in gravitoelectromagnetism.  Imagine a test particle following a
circular orbit in the Schwarzschild geometry.  Suppose that the constant ``radius'' of the orbit is given by the
Schwarzschild radial coordinate $r$.  Let $\rho$ be the corresponding isotropic radial coordinate,
$r=\rho[1+GM/(2c^2\rho)]^2$.  It is well known (see, e.g., [1]) that the orbital period in terms of $r$ is given by the
``Keplerian'' formula $2\pi(r^3/GM)^{1/2} = 2\pi (\rho^3/GM)^{1/2}\:[1+GM/(2c^2\rho)]^3$, which to
$O(c^{-2})$ reduces to our result with $\rho=a$.

\newpage

\noindent References\\
\baselineskip18pt
\vspace{.05in}
\begin{description}
\item{[1]} J.M. Cohen, B. Mashhoon, Phys. Lett. A 181 (1993) 353.

\item{[2]} B. Mashhoon, in:  Proceedings of the Workshop on the Scientific Applications of Clocks in Space, edited by
L. Maleki (JPL Publication 97-15, NASA, 1997), pp. 41-48.

\item{[3]} B. Mashhoon, F. Gronwald, D.S. Theiss, Ann. Physik 8 (1999) 135; gr-qc/9804008.

\item{[4]} W.B. Bonnor, B.R. Steadman, Class. Quantum Grav. 16 (1999) 1853.

\item{[5]} O. Semer\'{a}k, Class. Quantum Grav. 16 (1999) 3769.

\item{[6]} B. Mashhoon, N. O. Santos, Ann. Physik 9 (2000) 49; gr-qc/9807063.

\item{[7]} B. Mashhoon, F. Gronwald, H.I.M. Lichtenegger, in:  Gyros, Clocks, Interferometers...$\;$:  Testing
Relativistic Gravity in Space (Lecture Notes in Physics 562), edited by C. L\"{a}mmerzahl, C.W.F. Everitt and F.W.
Hehl, Springer, Berlin, 2001, pp. 83-108; gr-qc/9912027.

\item{[8]} D. Bini, R. T. Jantzen, B. Mashhoon, Class. Quantum Grav. 18 (2001) 670; gr-qc/0012065.

\item{[9]} R. Maartens, B. Mashhoon, D. Matravers, submitted for publication; gr-qc/0104049.

\item{[10]} D. Bini, R.T. Jantzen, B. Mashhoon, submitted for publication.

\item{[11]} F. Gronwald, E. Gruber, H. Lichtenegger, R. A. Puntigam, in:  Proc. Alpbach School on Fundamental Physics
in Space (ESA, SP-420, 1997), p. 29.

\item{[12]} L. Iorio, Int. J. Mod. Phys. D 10 (2001) 465; gr-qc/0007014.

\item{[13]} L. Iorio, Class. Quantum Grav. 18 (2001) 4303; gr-qc/0007057.

\item{[14]} H.I.M. Lichtenegger, F. Gronwald, B. Mashhoon, Adv. Space Res. 25 (2000) 1255; gr-qc/9808017. 

\item{[15]} A. Tartaglia, Class. Quantum Grav. 17 (2000) 783.

\item{[16]} A. Tartaglia, Class. Quantum Grav. 17 (2000) 2381.

\item{[17]} H.I.M. Lichtenegger, W. Hausleitner, F. Gronwald, B. Mashhoon, Adv. Space Res., in press; gr-qc/0101089.

\item{[18]} L. Iorio, H.I.M. Lichtenegger, B. Mashhoon, submitted for publication; gr-qc/0107002.

\item{[19]} I. Ciufolini, J. A. Wheeler, Gravitation and Inertia, Princeton University Press, Princeton, 1995.  

\item{[20]} B. Mashhoon, in:  Reference Frames and Gravitomagnetism, edited by J.-F. Pascual-S\'{a}nchez, L.
Flor\'{i}a, A. San Miguel and F. Vicente, World Scientific, Singapore,
2001, pp. 121--132; gr-qc/0011014. 

\item{[21]} B. Mashhoon, Astrophys. J. 223 (1978) 285.

\item{[22]} L.D. Landau, E.M. Lifshitz, The Classical Theory of Fields, Pergamon Press, Oxford, 1971, pp. 321-322. 

\end{description}
\vspace{.5in}

\noindent Figure captions\\
\begin{description}
\item{Figure 1.} Schematic plot of the unperturbed elliptical orbit.  The orbital plane depicted here coincides with
the plane of the osculating ellipse for the perturbed orbit at $t=0$.\\

\item{Figure 2.}  $\Delta{\cal T} = {\cal T}_{+} - {\cal T}_{-}$ in seconds versus the orbital eccentricity $e,
0\leq e\leq 0.95$, for a satellite in orbit around the Earth.  The common orbital parameters are $\varphi_0 =
45^{\circ}$ and
$g=-195^{\circ}$.  The solid line indicates an equatorial orbit $(i=0)$, while the broken line indicates an inclined
orbit with $i=45^{\circ}$.
\end{description}

\newpage
\epsfbox{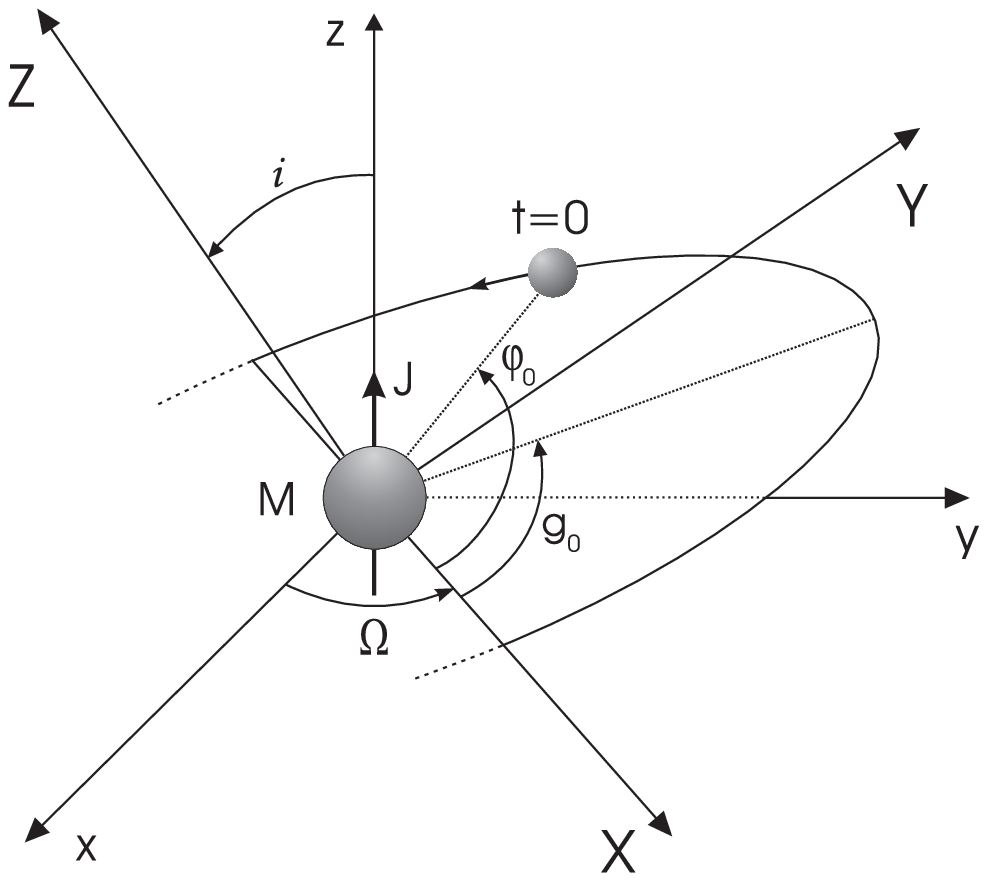}

\epsfbox{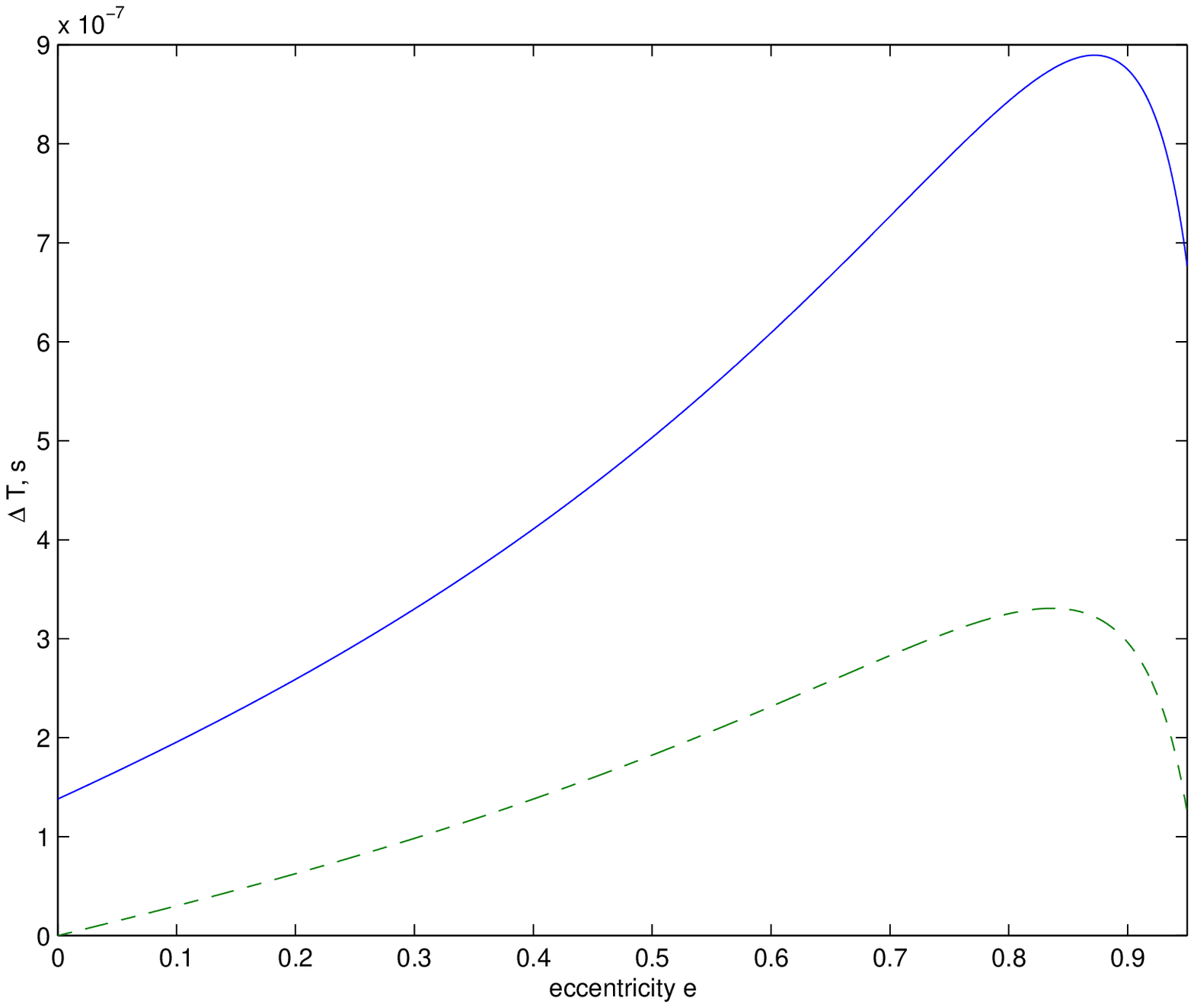}

\end{document}